\documentclass[utf8]{frontiersSCNS} 

\usepackage{url,hyperref,lineno,microtype,subcaption}
\usepackage[onehalfspacing]{setspace}
\usepackage{enumerate}
\usepackage{amsmath}
\usepackage{cases}
\usepackage{longtable}
\usepackage{hyperref}
\usepackage{epstopdf}
\usepackage{amsmath,bm}
\usepackage{amssymb}
\usepackage{natbib}
\usepackage{morefloats}
\usepackage{multirow}
\usepackage{array}
\usepackage{verbatim}
\usepackage{caption}
\usepackage{subcaption}

\newcommand\be{\begin{equation}}
\newcommand\ee{\end{equation}}

\newcommand*{\xu}{\color{black}}
\makeatletter
\newcommand{\@extraAuth}{}
\makeatother

\def\keyFont{\fontsize{8}{11}\helveticabold }
\def\firstAuthorLast{Lazarian, Xu, \& Hu} 
\def\Authors{Alex Lazarian$^{1,2,*}$,
Siyao Xu$^{3}$, Yue Hu$^{1}$}
\def\Address{$^{1}$Department of Astronomy, University of Wisconsin, 475 North Charter Street, Madison, WI 53706, USA \\
$^{2}$Centro de Investigación en Astronomía, Universidad Bernardo O’Higgins, Santiago, General Gana 1760, 8370993, Chile\\
$^{3}$Institute for Advanced Study, 1 Einstein Drive, Princeton, NJ 08540, USA; NASA Hubble Fellow}

\def\corrAuthor{Alex Lazarian}
\def\corrEmail{alazarian@facstaff.wisc.edu}

\begin{document}
\onecolumn
\firstpage{1}

\title[Cosmic rays in MHD turbulence]
{Cosmic ray propagation in turbulent magnetic fields} 

\author[\firstAuthorLast ]{\Authors} 
\address{} 
\correspondance{} 

\maketitle

\begin{abstract}
Propagation of cosmic rays (CRs) in 
turbulent and magnetized astrophysical media is a long-standing problem that requires both understanding of the properties of turbulent magnetic fields and their interaction with energetic particles. This review focuses on selected recent theoretical findings made based on the 
progress in understanding and simulating magnetohydrodynamic (MHD) turbulence. In particular, we address the problem of perpendicular and parallel propagation of CRs and identify the conditions when the perpendicular propagation is superdiffusive and diffusive. For the parallel diffusion, we discuss the problems of the traditionally used diffusion mechanism arising from pitch angle scattering and the possible solutions provided by the recently identified ``mirror diffusion" in the presence of turbulent magnetic mirrors. 

\tiny
 \keyFont{ \section{Keywords:} Cosmic rays, interstellar medium, MHD turbulence} 
\end{abstract}

\section{Introduction}
Nature demonstrates unique abilities to accelerate energetic particles. In astrophysical literature, such particles with energies ranging from {\xu $10^7$ eV to $10^{20}$ eV} are usually termed cosmic rays (CRs). The processes of acceleration include systematic First Order Fermi and stochastic Second Order Fermi acceleration mechanisms \citep{Mel69}, and shocks, magnetic reconnection, and turbulence are involved in the acceleration processes \citep{2011hea..book.....L}. CRs in the Galaxy have energy density and pressure comparable to those arising from interstellar turbulence and magnetic fields \citep{Park66}. Therefore, they are dynamically important.

{\xu The observed CR flux and (an)isotropy in their arrival directions depends on both their acceleration and propagation from their sources to the Earth \citep{Ginz64,Schlickeiser02}.} This review {\xu focuses on selected recent theoretical findings on CR propagation in turbulent magnetic fields, which are ubiquitous in space and astrophysical environments. CR propagation has numerous implications} on various processes in the solar wind, heliosphere, interstellar media, and intracluster media (see monograph by \cite{2011hea..book.....L} and references therein). 
{\xu It is a crucial part of CR acceleration process 
(e.g., \citealt{Jok82,Perri2009,LY14,Demi20,Morc20,2022ApJ...925...48X}). The diffusion in the vicinity of CR acceleration sites, e.g., supernova remnants, and pulsar wind nebulae, is important for understanding gamma-ray observations \citep{Dimau20,Xu21}.} Other implications include 
the solar modulation of Galactic CRs, space weather forecasting \citep{Par65,Jo71,SinH01},
the origin and chemical composition of CRs, ionization in molecular gas and circumstellar discs (e.g., \citealt{Schlk16,Padd18}), Fermi Bubble emission
\citep{Anjo20}, galactic winds (e.g., \citealt{Ipa75,Hol19,Hopk20,Quat22}), feedback heating in clusters of galaxies (e.g., \citealt{Guo08,Brun14}), modeling the synchrotron foreground emission for cosmic microwave background (CMB) radiation and redshifted $21$ cm radiation (e.g., \citealt{Chf02,Chh12}). 

In the propagation process, CRs interact with the pre-existing magnetohydrodynamic (MHD) fluctuations and the magnetic fluctuations created by themselves. The most notable {\xu example for the latter is the}  perturbations created by the streaming instability (see \citealt{Kulsrud_Pearce}). The suppression of streaming instability by {\xu various damping effects in the multi-phase interstellar medium 
\citep{Plot21,2022ApJ...925...48X,Samp23}} and by MHD turbulence \citep{YL02,FG04,La16} is an important effect that modifies the CR propagation (see \citealt{Krum20} and review by \citealt{2022FrP....10.2799L}).
Below we discuss the CR propagation in the pre-existing MHD turbulence. 

The interaction of CRs with magnetic turbulence has been investigated for decades 
\citep{Jokipii1966,Kulsrud_Pearce,SchlickeiserMiller,Giacalone_Jok1999},
mostly with ad hoc models of MHD turbulence, including superposition of MHD waves
\citep{Giacalone_Jok1999}, isotropic MHD turbulence (see \citealt{Schlickeiser02} and references therein), and {\xu 2D + slab superposition model for solar wind turbulence \citep{Mat90,Zank17}.} More recently, both MHD simulations with driven turbulence \citep{CV00,MG01,CLV_incomp,CL03,Bere14,KowL10} and solar wind observations (e.g., \citealt{Hor08,Luo10,For11}) show evidence for the statistical properties of MHD turbulence corresponding to theoretical expectations (\citealt{GS95,LV99}, see also \citealt{BL19} and references therein). {\xu As we discuss in the review, the progress in the understanding of MHD turbulence brings significant changes in understanding CR propagation.}
\footnote{For the sake of simplicity, we discuss only the so-called ``balanced" MHD turbulence, i.e., MHD turbulence with equal fluxes of Alfv\'en energy in opposite directions. This is not the case for Solar wind over distances less than 1 AU. However, the turbulence gets balanced at larger scales. The theory of imbalanced MHD turbulence is described in the book by \citealt{BL19}. }  

In what follows, in \S 2 we briefly discuss the basic properties and scaling relations of turbulent magnetic fields. \S 3 deals with CR superdiffusion and diffusion in the direction perpendicular to the mean magnetic field that arise from the perpendicular superdiffusion and diffusion of turbulent magnetic fields.
{\xu In \S 4, we discuss different mechanisms leading to the parallel diffusion of CRs, including a recently identified new mechanism, the mirror diffusion, that can significantly suppress the CR diffusion in the vicinity of CR sources. 
A summary is provided in \S 5.} 


\section{Properties of MHD turbulence that affect CR propagation}

\subsection{Compressible MHD turbulence and fast modes}

 We consider turbulence as a result of non-linear interactions inducing the energy cascade {\xu from large to small scales}. Therefore, other types of excitations, 
including the streaming instability (see \citealt{Kulsrud_Pearce,FG04,La16}), gyro-resonance instabilities (see \citealt{LB06}), are not classified as turbulence. The changes in {\xu plasma conditions} affect MHD turbulence and, through this, the CR propagation. For instance, partial ionization of astrophysical media can significantly affect the dynamics of CRs \citep{Xuc16}.

\begin{figure*}
	\centering
\includegraphics[width=.99\linewidth]{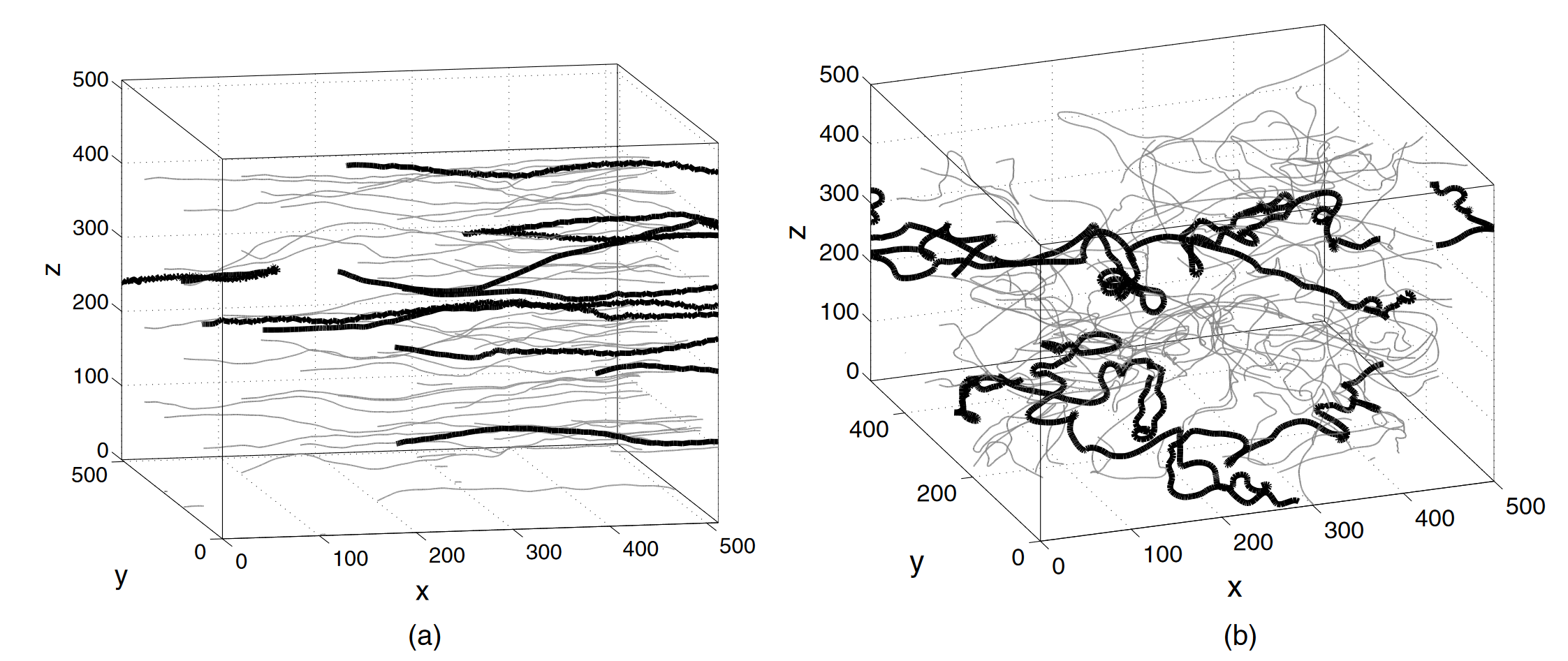}
	\caption{ Magnetic field lines (thin lines) and trajectories of CRs (thick lines) in (a) sub-Alfv\'enic turbulence with $M_A=0.3$ and (b) super-Alfv\'enic turbulence with $M_A=1.5$. From \citet{XY13}.}
\label{fig:3}
\end{figure*}

\begin{figure*}
	\centering
	\includegraphics[width=0.45\linewidth]{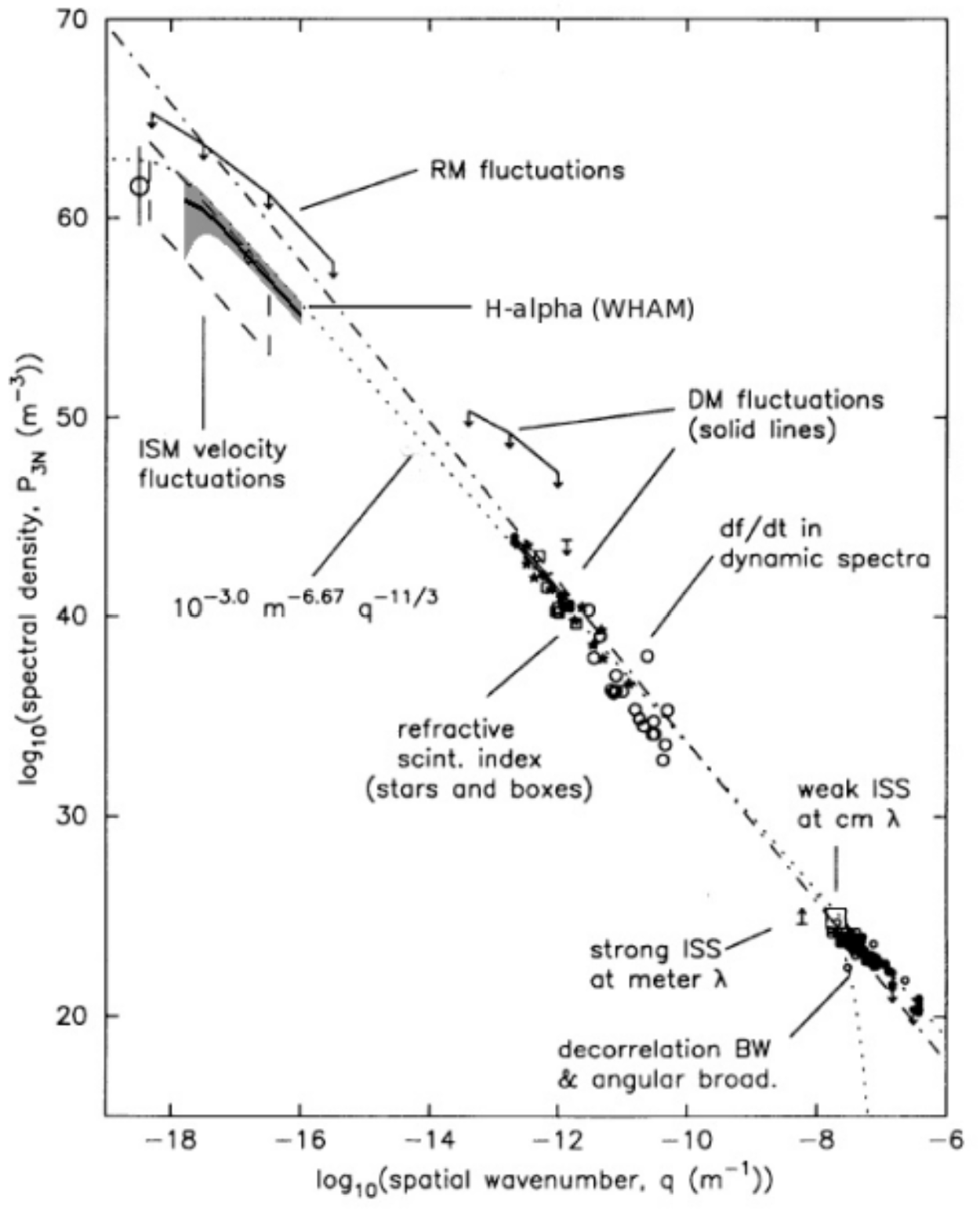}
 \includegraphics[width=0.45\linewidth]{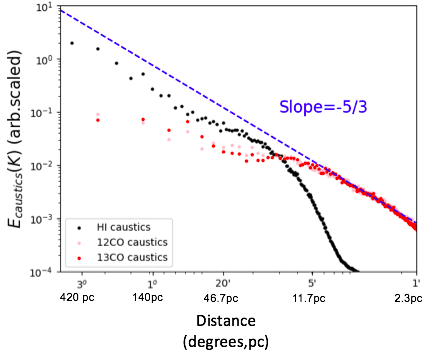}
	\caption{{\it Left panel:} ''Extended Big Power Law in the Sky'' of Galactic electron density fluctuations obtained by combining the scattering measurements of nearby pulsars in \citet{Armstrong95} and H$_\alpha$ measurements from WHAM data in \citet{CheL10}. {\it Right panel:} Power law of {\it velocity} fluctuations measured toward the Taurus molecular cloud. {\xu The horizontal axis corresponds to $\log_{10}$(spatial wavenumber) and the slope in the right panel is consistent with that in the left panel.} From \citet{2022arXiv220413760Y}.}
\label{fig:1}
\end{figure*}

The turbulence in magnetized media in typical astrophysical settings is injected at a large scale $L$, i.e., the injection scale, with the {\xu injected turbulent velocity at $L$ equal to $V_L$. The values of $L$ and $V_L$ depend on the driving mechanism of turbulence. For instance, for the interstellar turbulence driven by supernova explosions, the typical values of $L$ and $V_L$ are $\sim 100$ pc and $\sim 10$ km s$^{-1}$ \citep{ChaSh20}. We note that, unlike the bulk fluid velocity, the turbulent velocity is the averaged velocity difference between two points in space separated by a given length scale.} {\xu For MHD turbulence,} if the injection of turbulence happens with $V_L>V_A$, where $V_A$ is the Alfv\'{e}n speed, this is the case of super-Alfv\'{e}nic turbulence \footnote{\xu Super-Alfv\'{e}nic turbulence should be distinguished from super-Alfv\'{e}nic (e.g., solar wind) flow. In super-Alfv\'{e}nic turbulence, it is the averaged relative velocity at $L$, rather than the bulk flow velocity, that exceeds $V_A$. }.
If $V_L<V_A$, the turbulence is sub-Alfv\'{e}nic. The ratio $M_A=V_L/V_A$ is the Alfv\'{e}n Mach number, which is an important parameter characterizing the properties of MHD turbulence. The special case with $M_A=1$ corresponds to the trans-Alfv\'{e}nic turbulence, {\xu with energy equipartition between turbulence and magnetic fields at $L$. The trans-Alfv\'{e}nic turbulence was originally studied in \citet{GS95}. Other turbulence regimes were covered in \citet{LV99}.
Fig. \ref{fig:3} illustrates the different magnetic field structures in sub- and super-Alfv\'{e}nic turbulence.} The turbulent energy then cascades down to  a small scale $l_d$ that in most cases is determined by microphysical processes. 
The range of scales between $L$ and $l_d$ is the inertial range, and it is usually a very extended range in typical astrophysical settings. For instance, Figure \ref{fig:1} illustrates the Galactic electron density spectrum \citep{CheL10} induced by the interstellar turbulence in the warm ionized medium, and the velocity spectrum towards Taurus molecular cloud \citep{2022arXiv220413760Y}. 
The properties of turbulence over the inertial range are {\xu the most relevant} for the CR physics that we deal with in this review. {\xu All the scaling relations that we will discuss are within the inertial range of turbulence.}

The current understanding of MHD turbulence differs significantly from the earlier models adopted for CR studies in the literature (e.g., \citealt{Mat90,Kota_Jok2000,Qin2002}). 
MHD turbulence in a compressible medium (see \citealt{BL19}) can be approximated as a superposition of three cascades of fundamental modes, i.e., Alfv\'en, slow, and fast modes. {\xu The Alfv\'en modes play a dominant role in regulating the dynamics of MHD turbulence and shaping turbulence anisotropy. Unlike long-lived wave oscillations about a fixed equilibrium point in space, the reconnection diffusion of turbulent magnetic fields enables their turbulent motions in the direction perpendicular to the local magnetic field \citep{LV99}, with a limited lifetime determined by the turbulent eddy turnover time in the strong MHD turbulence regime.} 
Therefore, we use the term ``modes" rather than ``waves" for describing MHD turbulence. {\xu Another important property} of MHD turbulence is its scale-dependent anisotropy. \citet{1996JGR...101.7619M} reported {\xu the anisotropy being greater for smaller wavenumber.} By contrast, larger anisotropy toward smaller length scales (larger wavenumber) is seen in MHD turbulence simulations in the reference frame with respect to the local magnetic field \citep{CV00,CL02_PRL,CL03}. The latter is well expected by the MHD turbulence theory (\citealt{GS95}; henceforth GS95) based on the critical balance relation of Alfv\'en modes. \footnote{However, in the global reference frame with respect to the mean magnetic field adopted in \citet{GS95}, the scale-dependent anisotropy is not expected. The local reference frame, i.e., the only reference frame for one to see the scale-dependent anisotropy, was introduced in \citet{LV99}.}

The numerical decomposition of MHD turbulence into different modes has demonstrated that the interaction between fast modes and other modes, i.e., Alfv\'en and slow modes, is relatively weak for sub-Alfv\'enic  non-relativistic MHD turbulence \citep{CL02_PRL}. Therefore, the cascade of fast modes can be assumed independent of the cascades of Alfv\'en and slow modes. The cascade of fast modes is similar to the acoustic one in a high $\beta$ medium, where $\beta$ is the ratio of the plasma to magnetic pressure. In this regime, fast modes are mostly compressions of plasmas that propagate with the sound speed $c_s$. It is also shown by \citet{CL02_PRL} that in the opposite limit of a low $\beta$ medium, fast modes are expected to form a cascade similar to the acoustic one, even though the fluctuations are compressions of magnetic fields that propagate with $\sim V_A$. The numerical simulations by \citet{CL02_PRL,CL03} support that the cascade of fast {\xu modes} is very similar to the acoustic cascade for various values of $\beta$.

\begin{figure*}
	\centering	\includegraphics[width=.80\linewidth]{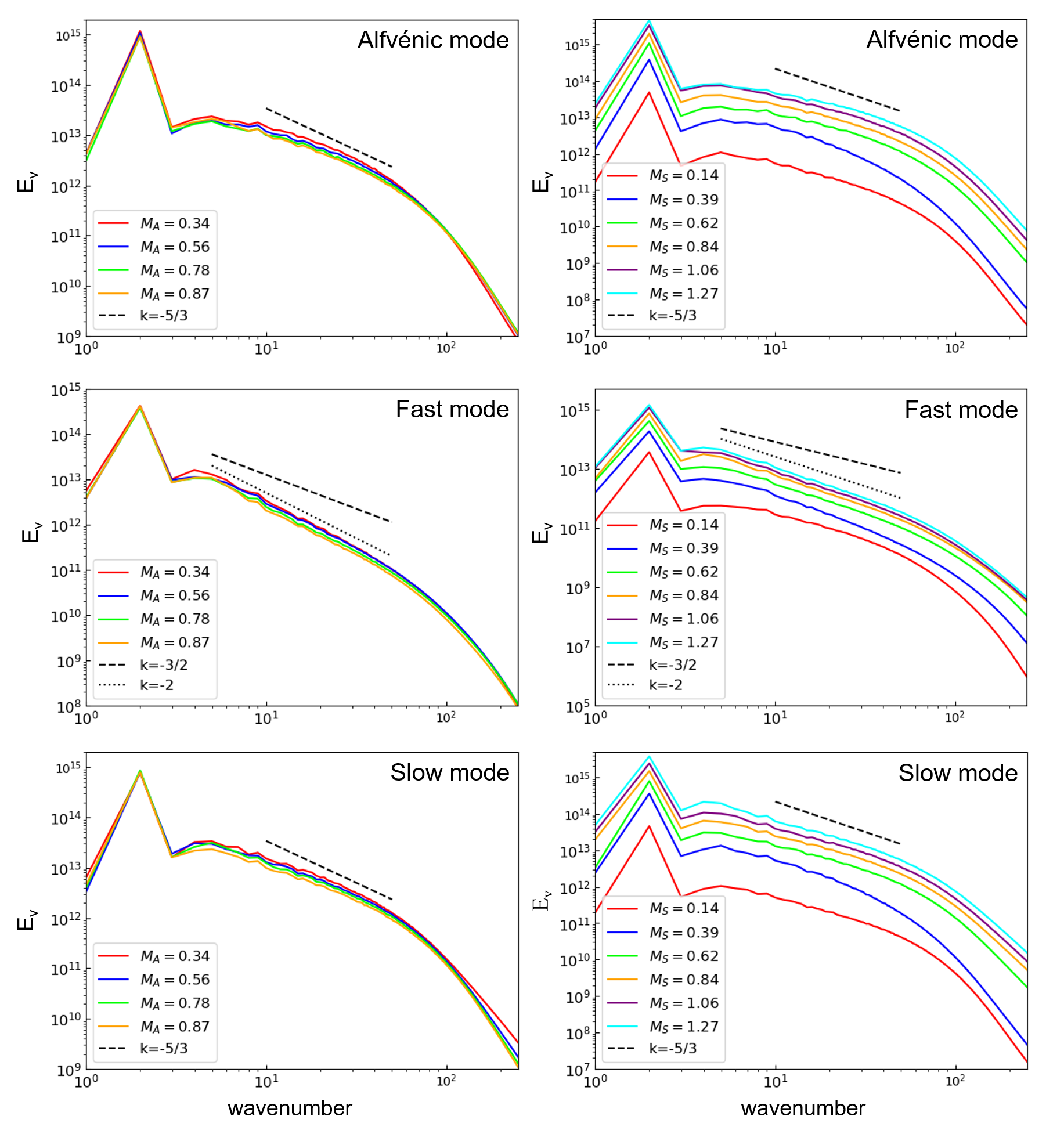}
	\caption{ \textbf{Left:} The turbulent kinetic energy spectra for Alfv\'en (top), fast (middle), and slow (bottom) modes for various $M_A$ values and $M_s\approx0.6$. \textbf{Right:} Same as left panels but for various $M_s$ values and $M_A\approx0.5$. 
 From \citet{2022MNRAS.512.2111H}. }
\label{fig:2}
\end{figure*}

The acoustic turbulence in non-magnetized gas has the energy spectrum
$E_s(k)\sim k^{-3/2}$ for low-amplitude perturbations. The spectrum tends to steepen with $E_s(k)\sim k^{-2}$ as the amplitude of perturbations increases. 
For fast modes in a high $\beta$ medium, the perturbation amplitude increases with increasing sonic Mach number $M_s = V_L /c_s$. For fast modes in a low-$\beta$ medium, it increases with the increase of $M_A$. The numerical results in \citet{CL03} are consistent with $E_f\sim k^{-3/2}$, while those in \citet{KowL10,2022MNRAS.512.2111H} are better fitted by $E_f\sim k^{-2}$
(see Fig.~\ref{fig:2}). The difference may be accounted for by appealing to the analogy with the acoustic turbulent cascade mentioned above. However, the issue has not been settled. In any case, similar to the acoustic cascade, fast modes have isotropic energy distribution. 

\subsection{Properties of Alfv\'en and slow modes}
{\xu Slow modes are cascaded by Alfv\'en modes,} and their anisotropy is imposed by that of Alfv\'en modes \citep{GS95,LG01,CL02_PRL,CL03}. Fig.~\ref{fig:2} shows the numerically measured turbulent kinetic energy spectra of decomposed MHD modes. As expected, the spectral scalings of Alfv\'en and slow modes in both compressible and incompressible media are similar \citep{CL03}. This allows appealing to high-resolution simulations of incompressible MHD turbulence \citep{Bere14} to test their spectral scalings.



{\xu The turbulent motions, rather than wave oscillations, of magnetic fields in Alfv\'en modes are enabled by the turbulent reconnection of magnetic fields
\citep{LV99} (henceforth LV99), which is an intrinsic part of MHD turbulence.} 
The LV99 theory treats Alfv\'en modes as a collection of turbulent eddies whose axes of rotation are aligned with the local magnetic field surrounding them. 
Due to turbulent reconnection, the fluid turbulent motions perpendicular to the {\it local} magnetic field are not constrained. As a result, an eddy with scale $l_\bot$ perpendicular to the local magnetic field induces Alfv\'{e}nic perturbation of scale $l_\|$ that propagates along the magnetic field with speed $V_A$. The timescale of this perturbation $l_\|/V_A$ should be equal to the eddy turnover time $l_\perp/v_l$. The corresponding relation between the parallel and perpendicular sizes of the eddy:
\begin{equation}
\frac{l_\|}{V_A}\approx \frac{l_\perp}{v_l},
\label{critbal}
\end{equation}
constitutes the modern understanding of the critical balance for Alfv\'enic turbulence (GS95). It follows that the GS95 theory is only valid in the so-called {\it local system of reference} defined by the local mean magnetic field of the eddies. This fact is confirmed by numerical simulations \citep{CV00,MG01,CLV_incomp}. The direction of the local magnetic field at a given length scale can differ significantly from the global mean magnetic field direction resulting from the large-scale averaging. For CR propagation, it is important that the local small-scale magnetic field is the magnetic field sampled by CRs. This justifies the perturbative approach for describing the propagation of low-energy CRs, even though the large-scale variations of the magnetic field can be significant. 

As the turbulent reconnection allows the turbulent cascade in the direction perpendicular to the local magnetic field, this cascade is not affected by the back-reaction of the magnetic field and remains Kolmogorov-like. For trans-Alfv\'{e}nic turbulence with $V_L = V_A$, this means that:
\begin{equation}
v_l\approx V_A \left(\frac{l_\bot}{L}\right)^{1/3},
\label{GS95perp}
\end{equation}
where $v_l$ is the turbulent velocity at $l_\perp$, and $l_\perp$ is the perpendicular size of a turbulent eddy. By combining Eq.~\eqref{GS95perp} and Eq. \eqref{critbal}, it is easy to obtain the scale-dependent anisotropy of trans-Alfv\'{e}nic turbulence in the {\it local system of reference}:
\begin{equation}
l_{\|}\approx L \left(\frac{l_\perp}{L}\right)^{2/3}.
\label{ll}
\end{equation}
From Eq. (\ref{ll}), it is evident that smaller eddies are more elongated along the local magnetic field. We note, however, that Eqs.  \eqref{GS95perp} and \eqref{ll} should be understood in the statistical sense. They represent the scaling relations between the most probable values of the quantities involved rather than the properties of individual eddies. The eddies have a distribution of scales. Using MHD turbulence simulations, \citet{CLV_incomp} provided an analytical fit to the probability distribution function of $l_\perp$ at a given $l_\|$. This distribution was adopted later in \citet{YL02,YL04} as well as the subsequent studies on CR scattering.

In super-Alfv\'enic turbulence, the magnetic field is of marginal importance at scales near $L$. Therefore, super-Alfv\'{e}nic turbulence has an isotropic Kolmogorov energy spectrum at large scales. However, as the turbulent velocity decreases along the energy cascade, i.e. $v_l\sim V_L (l/L)^{1/3}$, the effect of the magnetic field becomes more and more manifested. At the Alfv\'enic scale
\citep{Lazarian06}
\begin{equation}
l_A=L M_A^{-3},
\label{eq:A12}
\end{equation}
$v_l$ becomes equal to $V_A$, and the turbulence becomes fully magnetohydrodynamic. Its properties at scales smaller than $l_A$ can be described by trans-Alfv\'enic scaling, provided that $L$ in Eqs. \eqref{GS95perp} and \eqref{ll} is replaced by $l_A$. 

In sub-Alfv\'enic turbulence, it was shown in LV99 that below $L$, there is a range of scales where the cascade is in the {\it weak} turbulence regime. 
In this regime, the parallel scale of wave packets remains unchanged, i.e., $l_\|=L$, and {\xu oppositely directed wave packets} have to interact multiple times to get cascaded. The scaling obtained in LV99 for the weak turbulence\footnote{Weak turbulence is weak in terms of the non-linear interactions of oppositely directed wave packets.} under the assumption of the isotropic turbulence driving at $L$ is
\begin{equation}
v_l\approx V_L \left(\frac{l_\perp}{L}\right)^{1/2},
\end{equation}
and this result was supported by the subsequent analytical study by \citet{Gal00}. 
With the decrease of $l_\perp$, the intensity of interactions of Alfv\'{e}nic perturbations increases despite the decrease in the turbulence amplitude. It was shown in LV99 that at the transition scale
\begin{equation}
l_\text{tran}\approx L M_A^2,
\label{ltrans}
\end{equation} 
where $M_A<1$, the turbulence gets into the strong turbulence regime. For the sub-Alfv\'{e}nic turbulence at $l<l_\text{tran}$, there are 
\begin{equation}
v_l\approx V_L \left(\frac{l_\bot}{L}\right)^{1/3} M_A^{1/3},
\label{vAlf}
\end{equation}
and:
\begin{equation}
l_{\|}\approx L \left(\frac{l_\bot}{L}\right)^{2/3} M_A^{-4/3}.
\label{lpar}
\end{equation}
The above relations derived in LV99 differ from Eqs. \eqref{GS95perp} and \eqref{ll} for trans-Alfv\'{e}nic turbulence by the additional dependence on $M_A$.

{\xu To unify the formalism of scalings of super- and sub-Alfv\'enic turbulence,}
\citet{LX21} (henceforth LX21) proposed to introduce an effective injection scale: 
\begin{equation}
    L_\text{eff}=L M_A^{-4}.
    \label{eq:A11}
\end{equation}
Using $L_\text{eff}$, Eqs. \eqref{vAlf} and \eqref{lpar} can be rewritten as:
\begin{equation}
v_l\approx V_A \left(\frac{l_\bot}{L_\text{eff}}\right)^{1/3},
\label{vAlf1}
\end{equation}
and:
\begin{equation}
l_{\|}\approx L_\text{eff} \left(\frac{l_\bot}{L_\text{eff}}\right)^{2/3},
\label{lpar1}
\end{equation}
which take the same forms as Eqs. \eqref{GS95perp} and \eqref{ll} with $L$ replaced by $L_\text{eff}$. 
We note that unlike $l_A$ for super-Alfv\'enic turbulence, $L_\text{eff}$ does not have a particular physical meaning. 
LX21 introduced $L_\text{eff}$ to present the turbulent scaling on scales less than $l_\text{tran}$ for sub-Alfv\'enic turbulence in a more convenient way.

\subsection{Interactions of CRs with MHD turbulence}
Without magnetic fields, CRs would propagate ballistically and {\xu easily escape from cosmic accelerators and diffuse astrophysical media.} It is MHD turbulence that induces pitch-angle scattering, stochastic acceleration, and spatial diffusion of CRs along the magnetic field
(e.g., \citealt{Chan00,YL02,YL04,Strong07,Brunetti_Laz,YL08,Ly12,Blasi12,XY13,XLb18,Sio20,Lem20,Kemp22}), as well as superdiffusion and diffusion perpendicular to the mean magnetic field \citep{YL08,LY14,XY13,Mai22,2022MNRAS.512.2111H}.

Traditionally, the transport of CRs is separated in two categories, the transport along a magnetic field and the transport perpendicular to the magnetic field. 
In general, parallel and perpendicular transport take place simultaneously. 
Magnetic fluctuations are assumed to be small and therefore the magnetic field direction approximately coincides with that of the mean magnetic field. This is a very idealized presentation of the actual CR propagation in realistic astrophysical environments. In reality, the magnetic fluctuations are comparable to the mean magnetic field for trans-Alfv\'enic turbulence and are larger than the mean field for super-Alfv\'enic turbulence. Moreover, LX21 demonstrated that the processes of parallel and perpendicular diffusion can be interdependent.  

The field line wandering causes the diffusion of CRs perpendicular to the mean magnetic field. This effect was discussed extensively in the literature \citep{Jokipii1966}. 
However, perpendicular diffusion only applies on scales larger than $L$. 
On scales less than $L$, LV99 showed that the magnetic field lines are superdiffusive and exhibit fast dispersion, {\xu which is also numerically demonstrated by \citet{LVC04,Bere13}.} This fast separation of magnetic field lines induces superdiffusive transport of CRs {\it perpendicular} to the mean magnetic field direction \citep{LY14}. As we will discuss further, the superdiffusion of CRs is an essential but frequently ignored aspect of CR propagation.

As CRs move along magnetic field lines, they interact with magnetic fluctuations, 
{\xu including resonant (e.g., scattering, transit-time damping) and non-resonant (e.g., mirroring) interactions.} These interactions affect the {\it parallel} transport of CRs. The well-known interaction of CRs with magnetic fluctuations 
is gyroresonant scattering (i.e., pitch-angle scattering). It occurs as magnetic fluctuations induce fluctuations of the electric field in the reference frame of CRs with a frequency equal to the Larmor frequency of CRs. All MHD modes induce pitch angle scattering. In addition, a special type of interaction is related to compressible fluctuations, i.e., slow and fast modes, and is termed Transit-time Damping (TTD) interaction \citep{Schlickeiser02,XLb18}. This sort of interaction is associated with the surfing of CRs on the front of an oblique compressible wave. It changes the CR momentum parallel to the local magnetic field.

{\xu In super-Alfv\'enic turbulence, when the parallel mean free path of CRs is larger than $l_A$, the tangling of field lines acts to confine CRs, with an effective mean free path equal to $l_A$ \citep{Brunetti_Laz}, and the CR diffusion becomes isotropic. Super-Alfv\'enic turbulence in molecular clouds and starburst galaxies plays an important role in determining the diffusion of CRs
\citep{Krum20,Xu21}.}

The new understanding of the statistics and dynamics of turbulent magnetic fields
can shed light on some long-standing problems and observational puzzles
(e.g., \citealt{Palmer:1982,Evol14,Lop16,Krum20,Xu21}). It was found that the numerically tested model of MHD turbulence, rather than isotropic Kolmogorov turbulence, can satisfactorily interpret the high-precision AMS-02
measurements of CRs \citep{Forn21}. {\xu Updated multifrequency observations and direct CR measurements  (e.g., \citealt{Nav13,Orl18,Gabi19,Amat21}) 
request improved understanding on interactions of CRs with MHD turbulence, as well as measurements on the properties of astrophysical turbulent magnetic fields
\citep{Lazgt18,XuY21,HuX21,HuS22}.}

\section{CR propagation perpendicular to the mean magnetic field}
{\xu The perpendicular (super)diffusion of CRs arises from that of turbulent magnetic fields.} In the direction perpendicular to the mean magnetic field, the mean squared displacement $\langle y^2\rangle$ of CRs is proportional to $s^{\alpha}$, where $s$ is the distance traveled by CRs along the magnetic field.
$\alpha=1$ corresponds to normal diffusion, $\alpha > 1$ corresponds to superdiffusion, and $\alpha < 1$ corresponds to subdiffusion. {\xu Subdiffusion, i.e., compound diffusion, was proposed in, e.g, \citet{Kota_Jok2000} under the assumption that CRs are able to retrace the same magnetic field line. This is not possible due to the stochasticity of turbulent magnetic fields \citep{YL08,LY14,LX21}. Subdiffusion of CRs is not observed in MHD turbulence simulations 
\citep{XY13,2022MNRAS.512.2111H}.}

\subsection{Perpendicular superdiffusion of CRs}
{\xu Based on the kinetic theory, a CR particle is approximately shifted one Larmor radius perpendicular to the magnetic field during one scattering mean free path $\lambda_\|$
\citep{Jok87}.
The resulting perpendicular diffusion is usually negligible with $r_L \ll \lambda_\|$, where $r_L$ is the Larmor radius}.
Even in the limited case of strong scattering, i.e. in the case of the so-called Bohm diffusion, the perpendicular diffusion coefficient does not exceed $r_L c$, where $c$ is the speed of light. When the scattering is less efficient, CRs follow magnetic field lines during the time between the scattering events. As a result, the CR dynamics are very much affected by the stochasticity and (super)diffusion of turbulent magnetic fields. On scales smaller than $L$, the superdiffusion of magnetic fields was identified in LV99, {\xu which is a natural consequence of turbulent energy cascade and reconnection diffusion of magnetic fields
\citep{Laz14r}.}

The LV99 scaling for magnetic field {\xu superdiffusion} can be obtained by considering the {\xu energy cascade of Alfv\'en modes}. It is natural to assume that when one follows the magnetic field line over the parallel size $l_\|$ of a turbulent eddy, the magnetic field line undergoes perpendicular displacement equal to the transverse size $l_{\bot}$ of the eddy. The displacement $l_\bot$ can be either positive or negative, and therefore the dispersion $\langle y^2 \rangle$ of magnetic field lines increases in a random-walk manner with
\begin{equation}\label{y2}
d\langle y^2 \rangle\approx l_\bot^2 \frac{ds}{l_\|},
\end{equation}
where $ds$ is the distance measured along the magnetic field line, and the $\langle ...\rangle$ denotes an ensemble average. Using the above relation and the scaling relation between $l_{\|}$ and $l_{\bot}$ for sub-Alfv\'{e}nic turbulence given by Eq. (\ref{lpar}) and associating $l_\bot^2$ with $\langle y^2\rangle$, one gets (LV99):
\begin{equation}
\label{y3}
d\langle y^2 \rangle\approx \langle y^2 \rangle^{2/3} M_A^{4/3} L^{-1/3} ds,
\end{equation}
which indicates an accelerated separation of field lines. 
Superdiffusion of magnetic fields in MHD turbulence is analogous to Richardson dispersion \citep{Rich26} in 
hydrodynamic turbulence.
Numerically, the superdiffusion of turbulent magnetic fields was demonstrated in \citet{Laz04,Bere13}. 
{\xu The accelerated separation of magnetic field lines manifests that the ``frozen-in" condition is grossly violated in turbulent media.
More discussion on the relation between superdiffusion and flux-freezing breakdown 
can be found in \citet{Eyink2011}.
The latter has been numerically confirmed by \citet{Eyin13}.}

The essence of superdiffusion is easy to understand. As one follows neighboring magnetic field lines over the distance $s$, 
the divergence rate of field lines  increases as larger and larger turbulent eddies contribute to the dispersion of field line separations.
In terms of CRs dynamics, Eq. \eqref{y3} also applies to the dispersion of separations of CRs that move ballistically along magnetic field lines. The resulting perpendicular divergence of CR trajectories was  studied in \citet{YL08,LY14}, with the analytical predictions numerically confirmed in \citet{XY13}.
{\xu The dispersion of CR separations in the direction perpendicular to the mean magnetic field} in sub-Alfv\'enic turbulence is:
\begin{equation}
\langle y_{cr}^2\rangle \approx \frac{s^3}{L} M_A^4, ~~~M_A<1, ~s < \lambda_\|,
~ s< L,
\label{perpen1}
\end{equation}
where $s$ is the distance traveled by CRs along the magnetic field line, and $\lambda_\|$ is the CR parallel mean free path. 
In super-Alfv\'enic turbulence, there is 
 \begin{equation}
\langle y_{cr}^2\rangle \approx \frac{s^3}{L} M_A^3, ~~~M_A>1, ~s<\lambda_\|, ~ s<l_A.
\label{perpen2}
\end{equation}



{\xu In the case with efficient scattering,} the arguments above can be generalized to describe the {\xu perpendicular superdiffusion} of CRs while they propagate diffusively along the field lines. 
We have 
\begin{equation}
    s^2\approx D_\| t\approx\lambda_{\|} v_{cr} t,
\end{equation}
where $v_{cr}$ is the CR velocity, and $D_\|$ is the parallel diffusion coefficient. 
The average angle of a given magnetic field line with respect to the mean magnetic field is approximately 
\begin{equation}
    \alpha (s)\approx \frac{\langle y^2 (s) \rangle^{1/2}}{s}\approx \frac{s^{1/2}}{L^{1/2}} M_A^2,
    \label{alpha}
\end{equation}
where Eq. (\ref{perpen1}) is used.
It is easy to see that in the system of reference with respect to the mean magnetic field, the perpendicular displacement of CRs is $\lambda_{cr, \bot}\approx \alpha (s) \lambda_\|$ as they move over $\lambda_\|$ along the mean magnetic field. The perpendicular to the mean magnetic field component of CR velocity is $v_{cr, \bot}\approx \alpha (s) v_{cr, \|}$, where $v_{cr,\|}$ is the CR velocity along the mean magnetic field. 
The perpendicular propagation is a random walk with the step size $\lambda_{cr, \perp}$ and the velocity $v_{cr,\perp}$  that increase with $s$. Therefore, for a given $s$ the perpendicular diffusion coefficient is $\sim v_{cr, \bot} \lambda_{cr, \bot}\sim \alpha^2 (s) D_\|$. 
The motion of CRs perpendicular to the mean magnetic field can be described as  
\begin{equation}
\langle y^2_{cr} \rangle \approx \alpha^2 D_\| t\approx \frac{s^{3}}{L} M_A^4,~~~ ~~M_A<1,~~\lambda_\|\ll s, 
\label{perp_diff}
\end{equation}
It shows a similar superdiffusion as the case of inefficient scattering (Eq. \eqref{perpen1}).



We note that the concept of perpendicular superdiffusion of CRs contradicts some existing theories on small-scale CR transport, e.g. the Non-Linear Guiding Center Theory (NLGC) \citep{Matt03}. There it is assumed that in both parallel and perpendicular directions with respect to the mean magnetic field, the propagation of CRs is diffusive. 
{\xu However, test particle simulations in MHD turbulence support perpendicular superdiffusion of CRs on scales less than $L$
in both cases with ballistic and diffusive motions of CRs along magnetic field lines \citep{XY13,2022MNRAS.512.2111H}.}
\\



\subsection{Perpendicular diffusion of CRs}
The perpendicular diffusion of magnetic field lines in MHD turbulence takes place on scales larger than $l_A$ for super-Alfv\'enic turbulence and $l_\text{tran}$ for sub-Alfv\'enic turbulence (see Table 1 in \citealt{Lsup19}). For the former, magnetic field lines get entangled on the scale $l_A$ and this induces the random walk with the step $l_A$, i.e.:
\begin{equation}
    \langle y^2\rangle \approx l_A^2\frac{s}{l_A}=sl_A=sLM_A^{-3}, ~~M_A>1,~~s\gg l_A.
    \label{diff1}
\end{equation}
For sub-Alfv\'enic turbulence, when perpendicular scales are larger than $l_\text{tran}$, turbulence is in the weak turbulence regime, and the growth of magnetic field separation is diffusive with the step in the perpendicular direction $LM_A^2$ for every transposition along the magnetic field by $L$. Thus for $s\gg L$, there is
\begin{equation}
     \langle y^2\rangle \approx (LM_A^2)^2 \frac{s}{L}=sLM_A^4,~~M_A<1, ~~s\gg L.
     \label{diff2}
\end{equation}

For the CRs moving ballistically along magnetic field lines with $s<\lambda_\|$, the perpendicular diffusion of CRs is also given by Eqs. (\ref{diff1}) and (\ref{diff2}). If the CR propagation along the magnetic field is diffusive with $s\gg \lambda_\|$, 
{\xu the perpendicular diffusion has dependence on the parallel diffusion.}

In {\it super-Alfv\'enic turbulence} at scales $s\gg l_A$, the CR diffusion is isotropic. The characteristic diffusion coefficient is
\begin{equation}
    D\approx v_{cr} X,~~X=min[\lambda_\|, l_A].
\end{equation}

In {\it sub-Alfv\'enic turbulence}, {\xu if $L\ll s< \lambda_\|$, Eq. \eqref{diff2} leads to the perpendicular diffusion coefficient of CRs 
\citep{YL08}
\begin{equation}
   D_\perp \approx v_{cr} L M_A^4.
\end{equation}
If} $\lambda_\|\ll L \ll s$, {\xu CRs propagate diffusively over $L$ along the magnetic field line.} Below, we reproduce the derivation of the diffusion coefficient in Lazarian (2006).\footnote{The initial derivation was performed for the problem of thermal electron diffusion in clusters of galaxies. However, the physics of the diffusion of non-relativistic and relativistic particles are identical.} 
According to Eq. (\ref{diff2}) the magnetic field lines that guide the perpendicular diffusion undergo a random walk. As the magnetic field line is traced over $L$, its perpendicular transposition is $LM_A^2$. 
To cover the distance $\langle y_{cr}^2\rangle^{1/2}$ with the step size $LM_A^2$, it requires $N=(\langle y_{cr}^2\rangle^{1/2}/LM_A^2)^2$ steps. The time required for each step is $\delta t_{step}\approx L^2/D_\|$, and the total time is $\delta t \approx N \delta t_{step}$. Substituting this into the expression for the diffusion coefficient one gets
\begin{equation}
    D_{\bot}\approx \frac{\langle y_{cr}^2\rangle}{\delta t}\approx D_\| M_A^4.
    \label{diffperpnew}
\end{equation}
This agrees with the CR perpendicular diffusion coefficient in \citep{YL08}. 
{\xu This diffusion coefficient differs from that in the literature (see \citealt{Jokipii1966}) by having a dependence on $M_A^4$ rather than $M_A^2$. The $M_A^4$ dependence was numerically confirmed in  \citet{XY13}. }



\section{Propagation of CRs along the magnetic field}
\subsection{Gyroresonance scattering}
\label{ssec:gyroscat}
{\xu Gyroresonance scattering, i.e., pitch angle scattering, requires that the Doppler-shifted wave frequency is equal to the gyrofrequency of the particle or its harmonics.} 
Through the gyroresonant interaction with magnetic fluctuations, CRs experience diffusion in their pitch angles (the angle between the particle velocity and magnetic field direction) while moving along the local magnetic field. The distance that CRs travel along the magnetic field corresponding to the change of pitch angles by $90^\circ$ is $\lambda_\|$. 

For theoretical studies, the Quasilinear Theory (QLT) \citep{Jokipii1966} is frequently adopted in the literature. In QLT, particles are assumed to propagate along the magnetic field with infinitesimal fluctuations. Given $r_L \ll L$ and the turbulent energy cascade, the magnetic fluctuations that satisfy the gyroresonance condition can be sufficiently small for the above assumption to be valid. {\xu However, the QLT faces the so-called $90^\circ$ problem, with vanishing scattering close to $90^\circ$ and thus infinitely large $\lambda_\|$. The resonance broadening induced by, e.g., magnetic fluctuations, can help alleviate but not fully resolve the problem 
\citep{XLb18}.}


A remarkable consequence of the modern MHD turbulence theory is the inefficiency of gyroresonance scattering of CR with $r_L \ll L$ by Alfv\'en and slow modes. This effect was studied by \citet{Chan00,YL02,YL03}. \citet{YL02,YL03} identified fast modes as the main scattering agent of CRs with $r_L \ll L$ in MHD turbulence. \citet{YL02,YL03,XL20} adopted the anisotropic distribution of magnetic fluctuations measured from MHD turbulence simulations \citep{CLV_incomp}. This significantly shifted the paradigm of CR propagation and turbulent acceleration, as in earlier studies,  Alfv\'en modes were mostly considered as the source of gyroresonance scattering. 

The inefficiency of gyroresonance scattering by Alfv\'en and slow modes for CRs with $r_L \ll L$ arises from {\xu the scale-dependent anisotropy}. The gyroresonance condition requires that $r_L$ is comparable to $l_\|$. However, with scale-dependent anisotropy, there is $l_\bot\ll l_\|$ for small-scale magnetic fluctuations. 
As a result, a CR with $r_L\sim l_\|$ samples many uncorrelated turbulent eddies within one gyro-orbit. In addition, the energy cascade of Alfv\'en and slow modes is mainly in the direction perpendicular to the local magnetic field. With a steep parallel energy spectrum $E(k_\|)\sim k_\|^{-2}$ \citep{Bere15}, the power that induces the gyroresonance scattering falls relatively fast with the decrease of $r_L$. {\xu Both effects cause the inefficiency of gyroresonance scattering by Alfv\'en and slow modes \citep{Chan00,YL02,YL03,XL20}.}

{\xu As suggested by numerical simulations \citep{CL03},} fast modes have isotropic energy distribution. As a result, their interaction with CRs is not reduced due to geometrical factors. {\xu However, compared with Alfv\'en and slow modes, fast modes are more subject to damping effects because of their slower cascading rate.} The anisotropic nature of collisionless damping induces the preferential suppression of the modes with wavevectors perpendicular to the magnetic field, creating mostly slab-type structures of scattering fluctuations \citep{YL03,BL07}. This mitigates the supression of scattering by fast modes in the presence of significant damping \citep{YL03}. {\xu In a weakly ionized medium, fast modes are severely damped due to ion-neutral collisional damping \citep{Xuc16}. Only CRs with $r_L$ larger than the damping scale can be efficiently scattered. }
\\

\subsection{Transit-time Damping (TTD) interaction}
TTD interaction in plasma physics is usually associated with the damping of waves in collisionless plasmas. For CR studies, the TTD is the process of stochastically accelerating CRs by compressible MHD waves \citep{Schlickeiser02}. The process is easy to understand if one considers compressible waves oblique to the magnetic field direction. If the phase speed of waves is $v_\text{ph}$, and the angle between the wavefront and the magnetic field is $\alpha$, then the intersection point between the wavefront and the magnetic field moves with the speed $v_\text{inter}=v_\text{ph}/\sin\alpha\gg v_\text{ph}$ if $\alpha$ is small. If a CR moves along the magnetic field in the same direction with the speed close to $v_\text{inter}$, it can surf the wave, gaining or losing energy. {\xu With a linear resonance function,} the limitation of the process is that only waves in the limited range of oblique angles can interact with fast-moving CRs. {\xu Due to magnetic fluctuations and nonlinear decorrelation of turbulence, the resonance broadening effects in MHD turbulence plays an important role in determining the TTD efficiency by slow and fast modes beyond the threshold of the linear resonance \citep{XLb18}.}

{\xu The change of pitch angles caused by TTD is due to the particle acceleration.} TTD interaction increases the parallel component of CR momentum in a stochastic manner. {\xu Compared with gyroresonance scattering,} the advantage of TTD is that it is much less subject to damping effects that {\xu cut off the turbulent energy cascade.} 
Thus the process can act in settings where the gyroresonance scattering is inefficient \citep{YL03,BL07,Xuc16}. \\

\subsection{Mirror diffusion of CRs}
In addition to the pitch-angle scattering, it has long been known that CR propagation can also be affected by magnetic mirror reflection \citep{Fer49,Noe68,CesK73,Klep95,Chanmc00}. In particular, the magnetic mirroring effect was explored to resolve the $90^\circ$ problem that arises in the QLT describing the pitch-angle scattering (see Section \ref{ssec:gyroscat}). In these studies, the mirroring effect was invoked for trapping CRs that bounce back and forth between two mirror points. On the basis of improved understanding of MHD turbulence theory, \citet{LX21}  identified a new effect associated with CRs {\xu interacting stochastically with different mirrors}, which is termed {\it \xu mirror diffusion}.{\xu It serves as a new diffusion mechanism that can effectively confine CRs.} In MHD turbulence, compressions of magnetic fields, which arise from slow and fast modes in a compressible medium and {\xu pseudo Alfv\'en modes in an incompressible medium,}
naturally, induce the mirroring effect over a range of length scales. The properties of the magnetic mirrors are determined by the scaling properties of slow (pseudo Alfv\'en) and fast modes. Combined with the intrinsic perpendicular superdiffusion of turbulent magnetic fields arising from Alfv\'enic modes \citep{LV99,Eyin13,LY14}, 
this results in {\it mirror diffusion} of CRs along magnetic field lines. In other words, CRs are not trapped between mirrors, but exhibit a new type of diffusive propagation. 

\begin{figure}[htbp]
\centering   
\includegraphics[width=0.45\linewidth]{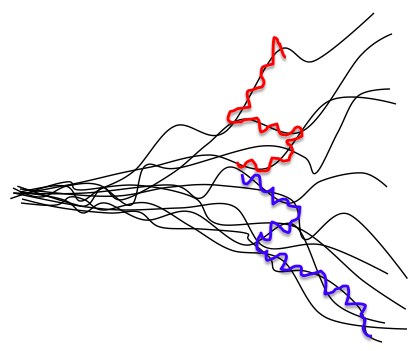}
 \includegraphics[width=0.45\linewidth]{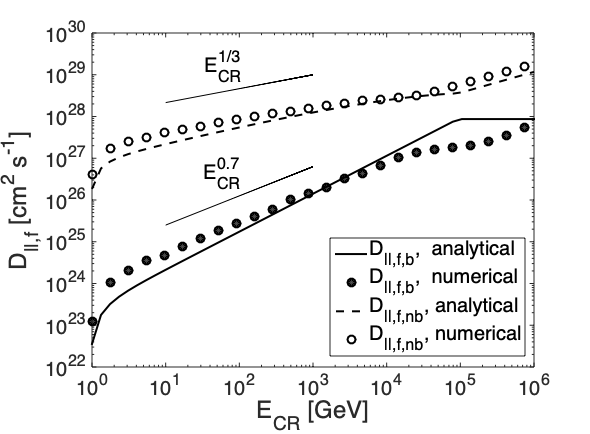}
\caption{{\bf Left:} Illustration of mirror diffusion. Thin lines represent turbulent 
magnetic field lines. Thick lines represent the trajectories of two CR particles whose initial separation is small. {\bf Right:} Parallel diffusion coefficients $D_{\|,f,b}$ of mirror diffusion 
and $D_{\|,f,nb}$ of scattering diffusion induced by fast modes. From LX21.}
\label{fig: untrfast}
\end{figure}

Magnetic compressions arising from slow and fast modes create magnetic mirrors that result in the reflecting of CRs. The mirroring effect caused by static magnetic bottles has been well studied in plasma physics (e.g., \citealt{Post58,Bud59,Noe68,Kulsrud_Pearce}). A CR with $r_L$ smaller than the variation scale of the magnetic field preserves its first adiabatic invariant, i.e. $p_{\bot}^2/B=$const. 
Therefore, as a particle moves along the magnetic field with the field strength increasing from $B_0$ to $B_0 + \delta b$,
its perpendicular momentum $p_\perp$ increases. 
It implies that 
%
with pitch angle cosine $\mu$
\begin{equation}\label{mu}
\mu < \mu_{lc}=\sqrt{\frac{\delta b}{B_0+\delta b}},
\end{equation}
{\xu the particle can be reflected at the mirror point, while particles with larger $\mu$'s, i.e., smaller pitch angles, can escape from the mirror.} 
Earlier studies considered that the magnetic mirrors trap the CRs until the gyroresonance scattering allows the particles to escape the mirrors \citep{CesK73}. CRs with $\mu<\mu_{lc}$ were considered ``trapped" in magnetic bottles and thus unable to diffuse. However, in \citet{LX21} it was shown that this is not true in realistic MHD turbulence. {\xu During the perpendicular superdiffusion, CRs cannot trace back the same magnetic field line. Instead, after each mirroring interaction, they always encounter a different mirror, leading to their diffusion along magnetic field lines (see Fig. \ref{fig: untrfast}).}




For mirroring interaction with fast modes, the corresponding parallel diffusion coefficient is (LX21)
\begin{subnumcases}
     { D_{\|,b} (\mu)=\label{eq: dufup}}
        v\mu k^{-1} = v L \Big(\frac{\delta B_f}{B_0}\Big)^{-4}  \mu^{9},  \nonumber
        ~~~~~~~~~\mu_\text{min,f} <\mu <\mu_{c} ,\\
        v\mu r_L, ~~~~~~~~~~~~~ \mu < \mu_\text{min,f},
\end{subnumcases}
{\xu where 
\begin{equation}
   \mu_c \approx  \bigg[ \frac{14}{\pi} \frac{ \delta B_f^2}{B_0^2} \Big(\frac{v}{ L \Omega}\Big)^\frac{1}{2} \bigg]^\frac{2}{11} 
\end{equation}
is the critical $\mu$ at the balance between pitch-angle scattering and mirroring, 
\begin{equation}
     \mu_\text{min,f}  = \sqrt{\frac{\delta B_f}{B_0}} \Big( \frac{r_L}{L} \Big)^{\frac{1}{8}},
\end{equation}
$v$ is the particle velocity, $\Omega$ is the gyrofrequency, $B_0$ is the mean magnetic field strength, and $\delta B_f$ is the magnetic fluctuation of fast modes at $L$.} For $\mu>\mu_c$ the diffusion is determined by scattering, and the scattering diffusion coefficient is typically much larger than that of mirror diffusion (see Figure \ref{fig: untrfast}). 
In the vicinity of CR sources, e.g., supernova remnants, the mirror diffusion of CRs with $\mu < \mu_c$ can prevent fast escape of CRs 
\citep{Xu21}. With pitch angle change by pitch-angle scattering, CRs undergo periods of slow mirror diffusion separated by periods of fast scattering diffusion, showing a L\'evy-flight-like characteristic.

\section{Summary}
The progress in understanding CR propagation was hindered for decades due to an inadequate understanding of MHD turbulence. {\xu Recent development in MHD turbulence theories, MHD turbulence simulations, as well as new observational techniques in measuring turbulence and magnetic fields, bring us new physical insight into the interaction of CRs with MHD turbulence and CR diffusion. The perpendicular (super)diffusion of CRs arises from that of turbulent magnetic fields regulated by Alfv\'en modes of MHD turbulence. The highly tangled magnetic field lines in super-Alfv\'enic turbulence provide additional confinement of CRs with an effective mean free path determined by the 
Alfv\'enic scale $l_A$. For parallel diffusion, the widely-used gyroresonance scattering, i.e., pitch-angle scattering, is inefficient in scattering CRs with large pitch angles and low-energy CRs in the presence of severe damping of fast modes, e.g., in a weakly ionized medium. With resonance broadening effects in MHD turbulence taken into account, the transit-time damping interaction can efficiently cause the change of pitch angles via stochastic particle acceleration. 
With both magnetic compressions and superdiffusion of magnetic field lines in MHD turbulence, CRs interacting with turbulent magnetic mirrors undergo mirror diffusion. It results in a slow diffusion of CRs in the vicinity of their sources and a L\'evy-flight-like propagation of CRs.

The perpendicular (super)diffusion and parallel diffusion of CRs strongly depend on the regime and properties of MHD turbulence. 
Combining the updated theoretical understanding of CR propagation and new observational techniques for mapping the characteristic parameters, e.g. $M_A$, in diverse astrophysical conditions 
(e.g., \citealt{Lazgt18,Zjf19,Humc19,Hugc20,XuY21,HuS22})
holds a great promise in realistic modeling of CR propagation and solving long-standing observational puzzles related to CRs. }


\section*{Conflict of Interest Statement}
The author declares that the research was conducted in the absence of any commercial or financial relationships that could be construed as a potential conflict of interest.

\section*{Author Contributions}
The article was devised and written by the authors of the article.
 
\section*{Funding}
 The research of AL and YH is supported by NASA ATP AAH7546. 
 SX acknowledges the support for this work provided by NASA through the NASA Hubble Fellowship grant \# HST-HF2-51473.001-A awarded by the Space Telescope Science Institute, which is operated by the Association of Universities for Research in Astronomy, Incorporated, under NASA contract NAS5-26555. 
 



\bibliographystyle{frontiersinSCNS_ENG_HUMS} 
\bibliography{xu}

\end{document}